\documentclass[preprint,12pt]{elsarticle}
\usepackage{amssymb,epsfig,graphicx,multirow,bm,dcolumn}
\journal{JMMM}
\begin{document}
\begin{frontmatter}

\title{Role of Tin and Carbon in the magnetic interactions in Mn$_3$SnC}
\author[gu]{V. N. Gaonkar, E. T. Dias}
\author[sinp]{Arka Bikash Dey}
\author[sinp]{Rajendra Prasad Giri}
\author[tifr]{A. K. Nigam}
\author[gu]{K. R. Priolkar \corref{krp}}\ead{krp@unigoa.ac.in}
\cortext[krp]{Corresponding author}
\address[gu]{Department of Physics, Goa University, Taleigao Plateau, Goa 403206 India}
\address[sinp]{Surface Physics and Material Science Division, Saha Institute of Nuclear Physics, 1/AF - Bidhannagar, Kolkata 700064, India }
\address[tifr]{Tata Institute of Fundamental Research, Dr. Homi Bhabha Road, Colaba, Mumbai 400005 India}

\begin{abstract}
In this paper we attempt to understand the role of tin and carbon in magnetic interactions in Mn$_3$SnC. Mn$_3$SnC exhibits a time dependent magnetic configuration and a complex magnetic ground state with both ferromagnetic and antiferromagnetic orders. Such a magnetic state is attributed to presence of distorted Mn$_6$C octahedra with long and short Mn--Mn bonds. Our studies show that C deficiency increases the tensile strain on the Mn$_6$C octahedra which elongates Mn--Mn bonds and strengthens ferromagnetic interactions while Sn deficiency tends to ease out the strain resulting in shorter as well as longer Mn--Mn bond distances in comparison with stoichiometric Mn$_3$SnC. Such a variation strengthens both, ferromagnetic and antiferromagnetic interactions. Thus the structural strain caused by both Sn and C is responsible for complex magnetic ground state of Mn$_3$SnC.
\end{abstract}
\date{\today}

\begin{keyword}
Antiperovskites, magnetostructural transformation, Mn$_3$SnC
\end{keyword}
\end{frontmatter}

\section{Introduction}
In recent times, Mn based antiperovskites have attracted considerable attention in terms of both fundamental research as well as potential technological applications.  A first order magnetostructural transition found in these compounds is responsible for many such properties like, large magnetocaloric effect (MCE) \cite{Tohei200394,Lewis200393,Aczel201490}, giant magnetoresistance (GMR) \cite{Kamishima200063,Li200572,Zhang2014115}, the invar effect, giant negative thermal expansion \cite{Wu2013114}, near zero temperature coefficient of resistance \cite{Chi2001120,Sun201062}, magnetostriction \cite{Shibayama2011109} and piezomagnetic effects \cite{Lukashev200878}.

Amongst these antiperovskites, Mn$_3$SnC exhibits a large magnetic entropy change near room temperature  ($\Delta S_{max} \sim $80.69 mJ/cm$^{3}$K$^{-1}$ under a magnetic field of 2T), comparable to those observed in contemporary magnetic refrigerant materials \cite{Wang200985}. Despite its relatively simple cubic structure, the compound transforms from a room temperature paramagnetic (PM) state to a high volume magnetically ordered state with a complicated spin arrangement consisting of antiferromagnetic (AFM) and ferromagnetic (FM) components via a spontaneous first order transition at 280K \cite{Dias201548}. The non collinearity of Mn spins in the transformed state has been attributed to a novel magnetic structure obtained from neutron diffraction studies. Unlike in the crystal structure, the magnetic unit cell ($a\sqrt{2}$, $a\sqrt{2}$, $a$) of Mn$_3$SnC generated using a propagation vector $k = \left [{1\over 2}, {1\over 2}, 0 \right ]$ consists of two types of Mn atoms. Two of the three Mn atoms present themselves in a square configuration with a net antiferromagnetic moment of 2.4$\mu_{B}$ per Mn in a plane perpendicular to $z-$ axis. While, the third Mn atom has a pure ferromagnetic moment of 0.65$ \pm $0.15$ \mu_{B} $ along the 001 direction \cite{Dias201548}. In contrast, Mn$_3$GaC, which also undergoes a first order transition has a collinear AFM structure described by a propagation vector $k = \left [\frac{1}{2}, \frac{1}{2}, \frac{1}{2} \right ]$ \cite{Fruchart19708,Cakir2014115}.

Dynamics of the first order transition in Mn$_3$SnC suggests that all three Mn spins order ferromagnetically along 001 direction before two of them align antiferromagnetically in a plane transverse to $c$ axis \cite{PhysRevB.96.014436}. Such a flipping of Mn spins causes mechanical strain which is in addition to the strain produced due to change in unit cell volume at the phase transition. Hence understanding the magnetic interactions between the Mn atoms and the role of Sn and C atoms in modifying these interactions becomes important.

Recent x-ray absorption fine structure (XAFS) studies on Mn$_3$GaC have shown the presence of distortions in Mn$_6$C octahedra which result in long and short Mn-Mn bond distances, while the Mn-C distance shows negligible variation across the magnetostructural transition \cite{Dias4996933}. This indicates that the magnetic interactions are largely due to Mn - Mn indirect interactions where the longer Mn-Mn bonds support ferromagnetic order and the shorter ones aid antiferromagnetic order. Though, distortions of the Mn$_6$C octahedra are also seen in Mn$_3$SnC, they are slightly different from those seen in Mn$_3$GaC and the differences are believed to be due to larger size of Sn as compared to Ga \cite{Dias201548}. This reiterates the importance of understanding the role of C and Sn atoms in magnetic interactions in Mn$_3$SnC. In order to understand the nature of magnetic interactions in Mn$_3$SnC and the role of Sn and C atoms in them, we have prepared two series of compounds, (i) with varying content of carbon (Mn$_3$SnC$_{1-x}$, $0\le x \le 0.2$) and (ii) with increasing tin deficiency (Mn$_3$Sn$_{1-y}$C, $0 \le y \le 0.15$). Since Sn and C have no direct role in the magnetism of Mn$_3$SnC, their presence (or absence) will affect the lattice strain and/or the magnetic interactions and thus will allow us to understand the magnetostructural coupling in such antiperovskite compounds.

\section{Experimental}
Polycrystalline Mn$_3$SnC$_{1-x}$ and Mn$_3$Sn$_{1-y}$C samples were prepared by taking together correctly weighed starting materials (Mn, Sn and C powders) and mixing them intimately in the appropriate molar ratio. About 15 \% excess graphite powder was added to compensate for a possible carbon deficiency during the reaction \cite{Lewis200393}. The resulting mixtures were then pressed into individual pellets and encapsulated each in a separate, evacuated quartz tube before sintering at 1073K for the first 48 hours and at 1150K for the next 120 hours. After quenching to room temperature, the product were powdered, mixed and reannealed  under the same conditions to obtain a homogeneous sample. To achieve carbon deficient samples, the total carbon content was reduced in steps of 5\%. For instance, the stoichiometric Mn$_3$SnC had 115\% by weight of carbon required and Mn$_3$SnC$_{0.8}$ was prepared with 95\% of carbon content required as per calculations. The final carbon content was estimated from CHNS analyzer and the obtained values are listed in table \ref{table1}. Room temperature x-ray diffraction (XRD) studies were carried out using a Rigaku diffractometer with CuK$_\alpha$ radiation to determine the phase formation and purity of the compound formed. Thermal expansion across the first order transition was measured from XRD patterns recorded in the temperature range of 25-300K using BL18B at Photon Factory, Japan. Temperature (5K - 350K) and field dependent (0 - 7T) magnetization measurements were performed using Quantum Design SQUID magnetometer. Magnetization measurements as a function of temperature were performed in an applied field of 100 Oe during zero field cooled (ZFC), field cooled cooling (FCC) and field cooled warming (FCW) cycles. For measurements as a function of field, compounds were first cooled in zero field to appropriate temperature and the field was ramped in the interval 0 - 7T.

\section{Results and Discussion}

\begin{table*}
\setlength{\tabcolsep}{1.5pc}
\caption{\label{table1} Elemental content, lattice parameter $a$ \AA and transition temperature ($T_{ms}$ (K)) obtained for Mn$_3$SnC$_{1-x}$, $x$ = 0, 0.05, 0.1, 0.15 and 0.2 and Mn$_3$Sn$_{1-y}$C, $y$ = 0.05, 0.1 and 0.15. Quantities in the parenthesis are uncertainties in the last digit.}
\begin{center}
\begin{tabular*}{\textwidth}{@{}lccccc}
\hline
Sample Name & \multicolumn{3}{c}{Element Content} & Lattice & Transition\\
 & Mn & Sn & C & Constant (\AA) & Temperature (K) \\
\hline
Mn$_3$SnC & 2.98(9) & 0.99(3) & 1.13(11) & 3.9938(1) & 284(1) \\
Mn$_3$SnC$_{0.95}$ & 3.06(9) & 0.97(3) & 1.13(12) & 3.9932(4) & 284(1)\\
Mn$_3$SnC$_{0.9}$ & 2.98(9) & 0.99(3) & 0.90(10) & 3.9942(3) & 283(1)\\
Mn$_3$SnC$_{0.85}$ & 2.98(9) & 0.99(3) & 0.82(9) & 3.9893(3) & 277(1)\\
Mn$_3$SnC$_{0.8}$ & 3.01(9) & 0.97(3) & 0.80(9) & 3.9888(5) & 270(1)\\
Mn$_3$Sn$_{0.95}$C & 2.98(9) & 0.93(3) & 0.96(10) & 3.9909(3) & 274(1)\\
Mn$_3$Sn$_{0.9}$C & 2.97(9) & 0.91(3) & 0.97(10) & 3.9896(4) & 272(1)\\
Mn$_3$Sn$_{0.85}$C & 2.99(9) & 0.87(3) & 0.95(10) & 3.9862(4) & 248(2)\\
\hline
\end{tabular*}
\end{center}
\end{table*}

\begin{figure}[htb]
\begin{center}
\includegraphics[width=\columnwidth]{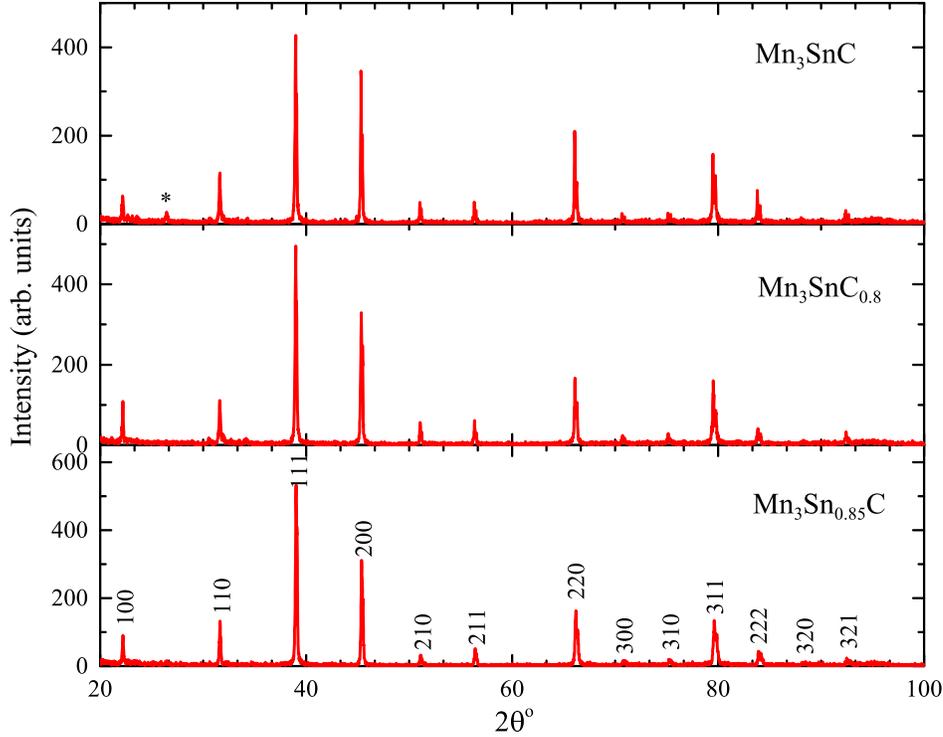}
\caption{Room temperature x-ray diffraction patterns for Mn$_3$SnC, Mn$_3$SnC$_{0.8}$ and Mn$_3$Sn$_{0.85}$C. * indicates graphite impurity phase.}
\label{fig:XRD}
\end{center}
\end{figure}

X-ray diffraction patterns were recorded on all prepared samples. All the compounds were found to be cubic crystallizing in {\em Pm-3m} space group. The patterns were Rietveld refined to calculate the lattice parameters as well as check for presence of any impurity phases. Minor impurity phase ($\sim$ 1\%) of graphite was detected in some of the patterns. The lattice parameters and the contents of Mn and Sn estimated form refinement of XRD patterns along with carbon content listed in Table \ref{table1}.  XRD patterns of three selected compositions, Mn$_3$SnC, Mn$_3$SnC$_{0.8}$ and Mn$_3$Sn$_{0.85}$, are presented in Fig. \ref{fig:XRD}. Lattice constant $a$, shows a general decrease with decreasing C and Sn content. While the change in $a$ is small with the decrease in C content, a much larger decrease in $a$ is seen as the Sn content is varied from 1 to 0.85. Such a variation, especially with C content is also seen in Mn$_3$GaC \cite{Dias2014363}.

Further in the paper, attention is focussed on three compositions, Mn$_3$SnC, Mn$_3$SnC$_{0.8}$ and Mn$_3$Sn$_{0.85}$. This is because, Mn$_3$SnC$_{0.8}$ and Mn$_3$Sn$_{0.85}$ show a maximum deviation in all properties from the stoichiometric compound, Mn$_3$SnC. This also agrees with the main objective of the paper which is to understand the role of C and Sn in magnetostructural interactions in Mn$_3$SnC.

\begin{figure}[htb]
\begin{center}
\includegraphics[width=\columnwidth]{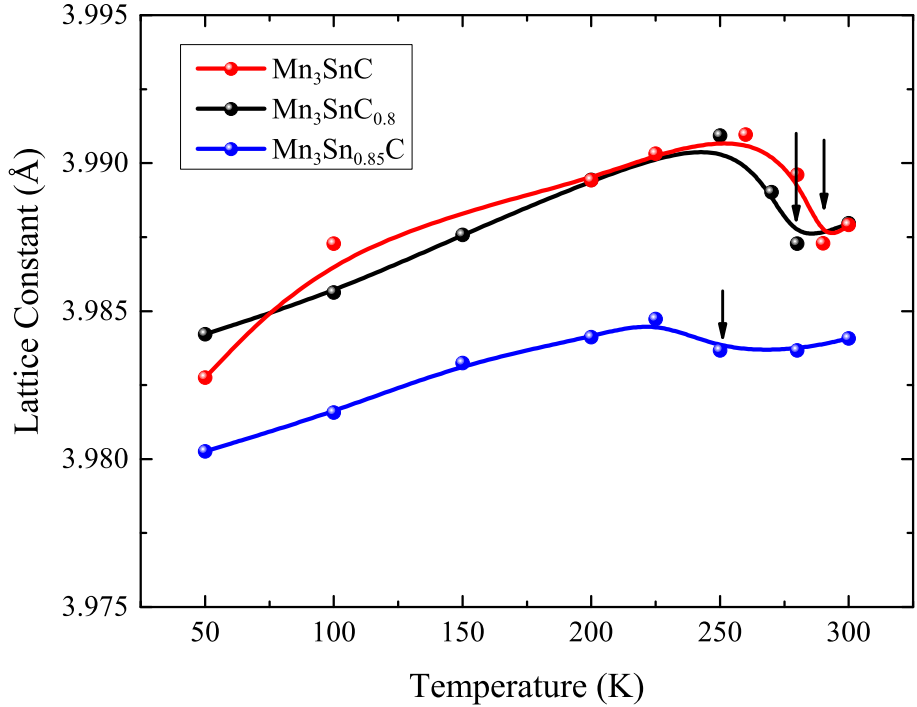}
\caption{Variation of lattice constant as a function of temperature in Mn$_3$SnC, Mn$_3$SnC$_{0.8}$ and Mn$_3$Sn$_{0.85}$C. Arrows indicate magnetostructural transition temperature.}
\label{fig:lattice}
\end{center}
\end{figure}

XRD patterns of Mn$_3$SnC, Mn$_3$SnC$_{0.8}$ and Mn$_3$Sn$_{0.85}$ recorded as a function of temperature were used to determine thermal evolution of lattice constants across the first order transition. The results can be seen in Fig. \ref{fig:lattice}. A clear increase in lattice constant at the magnetostructural transition temperature can be seen in Mn$_3$SnC and Mn$_3$SnC$_{0.8}$ compounds. The change in lattice volume is of the order of 0.1\% and is in agreement with the values reported for Mn$_3$SnC earlier cite{Dias201548}. The transition temperature noted from the discontinuity in lattice parameters agrees well with that obtained from magnetization measurements that are reported later. The values of transition temperature, obtained from magnetization data, are listed in Table \ref{table1} and indicate a decrease in transition temperature from 284K to 270K as the C content is reduced from 1.0 to 0.8. In the case of Mn$_3$Sn$_{0.85}$C both, the transition temperature and the \% lattice expansion at the transition are suppressed. In fact, the transition broadens considerably, extending from about 275K to 225K. Such a broad transition seen in Fig \ref{fig:lattice} is also visible in magnetization measurements on this compound.

\begin{figure}[htb]
\begin{center}
\includegraphics[width=\columnwidth]{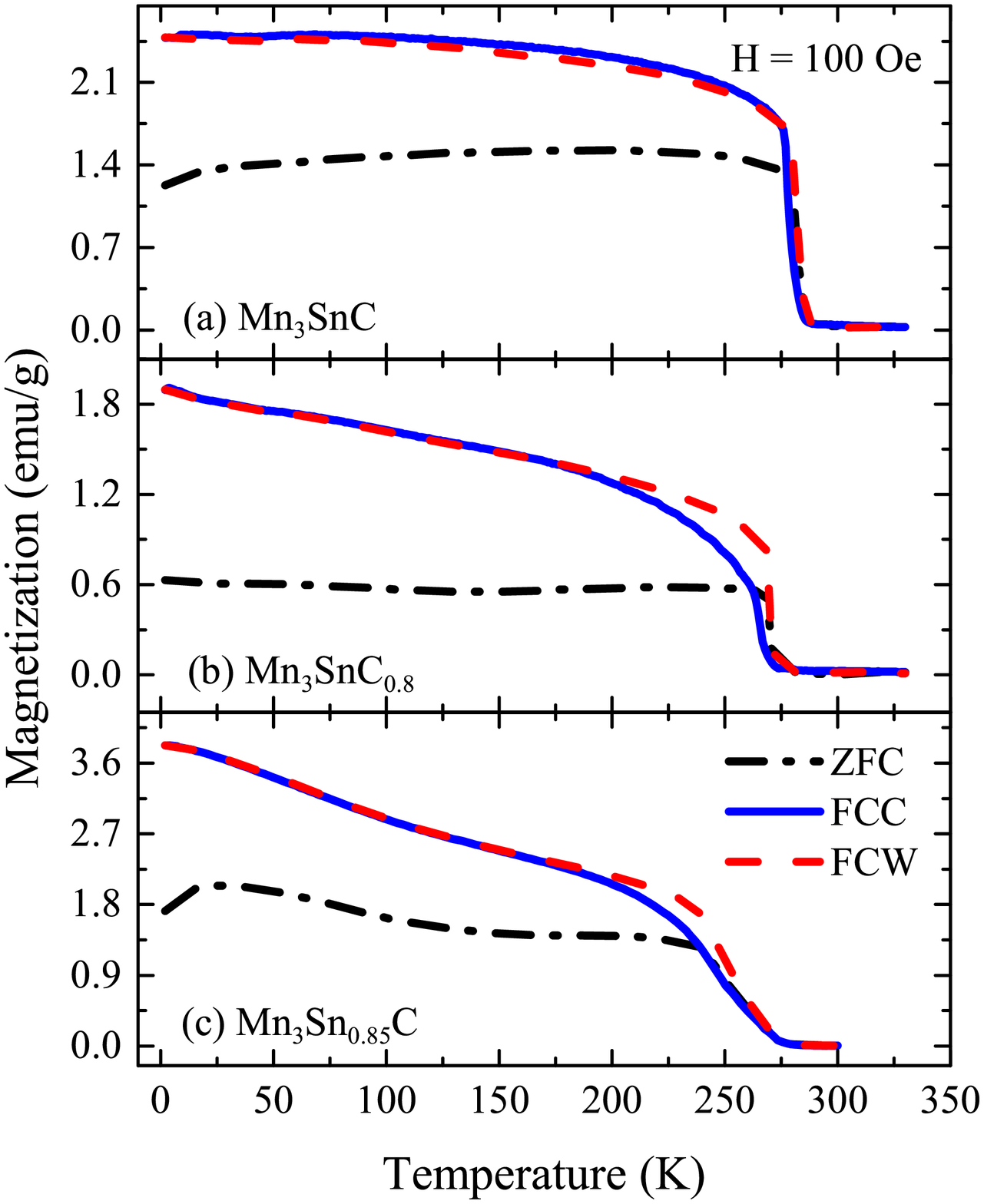}
\caption{Temperature dependent magnetization of Mn$_3$SnC, Mn$_3$SnC$_{0.8}$ and Mn$_3$Sn$_{0.85}$C recorded during ZFC, FCC and FCW cycles in 100 Oe.}
\label{fig:MT}
\end{center}
\end{figure}

Magnetization as a function of temperature was recorded for all the studied samples in an applied magnetic field of 100 Oe using zero field cooled (ZFC) and field cooled (FC) protocols. For ZFC data, the samples were cooled in zero applied field down to 5K and magnetization was recorded during warming in applied field. Magnetization was also recorded in subsequent cooling and warming cycles giving field cooled cooling (FCC) and field cooled warming (FCW) data respectively. The data obtained for three representative compounds, Mn$_3$SnC, Mn$_3$SnC$_{0.8}$ and Mn$_3$Sn$_{0.85}$C is presented in Fig. \ref{fig:MT}. For all three samples, magnetization shows a transition akin to paramagnetic to ferromagnetic transition at about the same temperature as the volume discontinuity observed in XRD studies indicating it to be a magnetostructural transition. Further, presence of hysteresis in the warming and cooling data indicated first order nature of the transition. Such a transition was also observed in the magnetization data of the other samples and the transition temperatures ($T_{ms}$) are listed in Table \ref{table1}.

From Fig. \ref{fig:MT} as well as from Table \ref{table1} it can be seen that, with decrease in both, C and Sn content, the transition temperatures decrease and this decrease is much sharper in case of Sn deficient compounds as compared to carbon deficient compounds. Another aspect  that is noticed is the broadness of the transition in Mn$_3$Sn$_{0.85}$C (Fig. \ref{fig:MT}(c)). At the transition, the increase in magnetization extends from 275K to 225K with the hysteresis between FCC and FCW cycles extending even further down to 200K. In comparison, in Mn$_3$SnC$_{0.8}$ (Fig. \ref{fig:MT}(b)), the increase in magnetization is much sharper but the hysteresis extends over a wide temperature range.

\begin{figure}[htb]
\begin{center}
\includegraphics[width=\columnwidth]{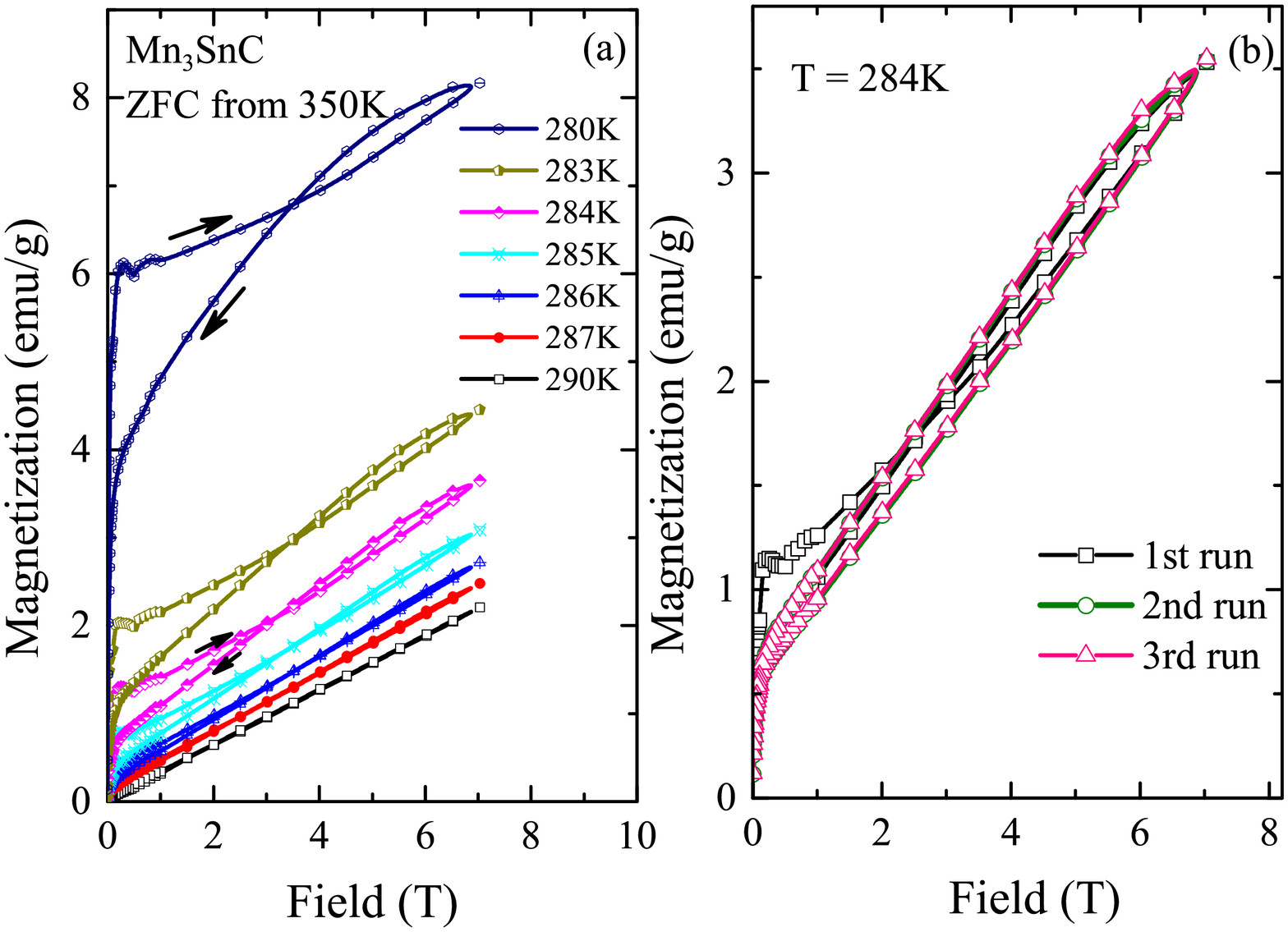}
\caption{(a) Zero field cooled two quadrant magnetic hysteresis loops in Mn$_3$SnC around $T_{ms}$ = 284K. (b) M(H) loop at $T_{ms}$ during first cycle after cooling in zero field from 350K and subsequent two cycles.}
\label{fig:MHC1}
\end{center}
\end{figure}

\begin{figure}[htb]
\begin{center}
\includegraphics[width=\columnwidth]{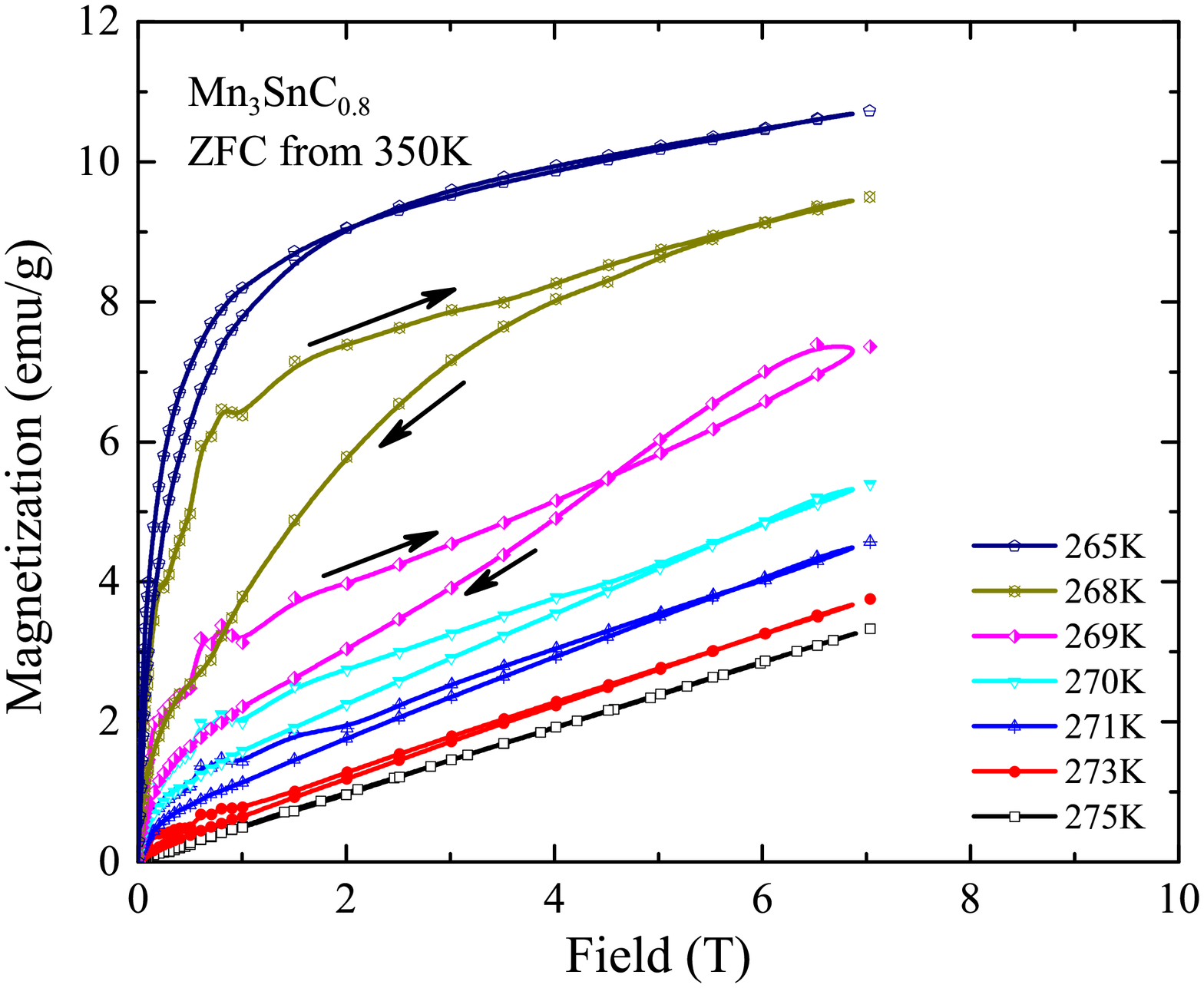}
\caption{Magnetic hysteresis loops in Mn$_3$SnC$_{0.8}$ around $T_{ms}$ = 270K. Each time the sample was cooled from 350K to target temperature in zero field.}
\label{fig:MHC08}
\end{center}
\end{figure}

\begin{figure}[htb]
\begin{center}
\includegraphics[width=\columnwidth]{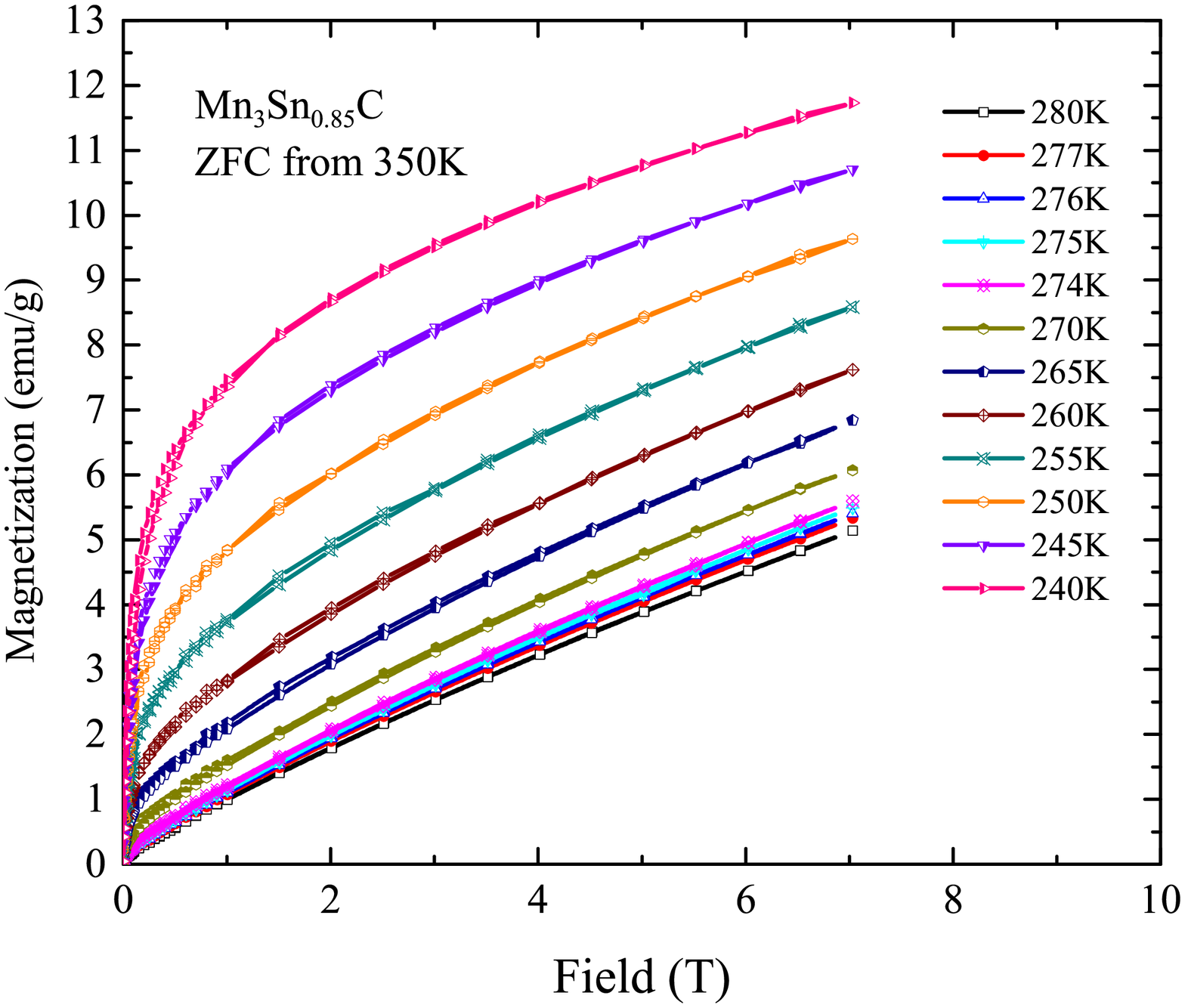}
\caption{Magnetization behavior in Mn$_3$Sn$_{0.85}$C around $T_{ms}$ = 245K during increasing and decreasing field cycles. The sample was cooled in zero field from 350K to the target temperature for every measurement.}
\label{fig:MHC085}
\end{center}
\end{figure}

Isothermal magnetization in a field interval of 0 to 7 Tesla have been recorded at several temperatures near the respective $T_{ms}$ of the three compounds. Magnetization was recorded in the first two quadrants, that is, while ramping the magnetic field up to 7T and then down to 0T. Before each measurement, the samples were cooled from 350K in zero applied field (ZFC) to the target temperature. At the transition temperature, magnetization was recorded by ramping the field up and down in the interval 0 to 7 Tesla for several cycles. All these results are presented in Figs. \ref{fig:MHC1}, \ref{fig:MHC08} and \ref{fig:MHC085} for Mn$_3$SnC, Mn$_3$SnC$_{0.8}$ and Mn$_3$Sn$_{0.85}$C respectively.   Magnetic ordering in Mn$_3$SnC is quite complex. Two of the three Mn atoms couple antiferromagnetically with their spins essentially pointing in the $x-y$ plane while the third Mn atoms align ferromagnetically with their spins pointing along the $z-$ axis. Such an ordering has been attributed to elongation of Mn$_6$C octahedra along one direction and shrinking along other two \cite{Dias201548}.

In case of Mn$_3$SnC (Fig. \ref{fig:MHC1}), magnetization behavior above the transition temperature (for instance at 290K) resembles that of a paramagnet. As the temperature is lowered and approaches $T_{ms}$, a crossover in magnetization curves is noticed. Here as the magnetic field is increased from 0 to 7T, the magnetization acquires a high value initially and then decreases slightly before increasing again up to 7T. While the field is decreased back to 0T, the magnetization has a normal behavior down to about 3T. Below this value of magnetic field there is a cross over and the magnetization loop is inverted which implies that the magnetization while decreasing the magnetic field has a lower value than that while increasing the field. This behavior continues well below the transition temperature. Upon reaching 0T, if the magnetic field is ramped again, magnetization follows a completely different route. There is no cross over and the loop is normal (see inset of Fig. \ref{fig:MHC1}). The inverted behavior is obtained only for the first cycle after cooling the sample in zero field from a temperature well above transition temperature (350K).

The inverted behavior of magnetization is also obtained in case of Mn$_3$SnC$_{0.8}$ (Fig. \ref{fig:MHC08}). Here, the magnetization loops are inverted over the entire field range except very near to transition temperature ($\sim$ 270K). The inverted hysteresis loops appear at 273K and are present down to 265K. A completely opposite behavior is noticed in case of Mn$_3$Sn$_{0.85}$C (Fig \ref{fig:MHC085}). All the hysteresis loops recorded in the temperature range from 270K to 240K exhibit normal behavior and show a monotonic increase with increase in applied field. Unlike in case of Mn$_3$SnC and Mn$_3$SnC$_{0.8}$, the magnetization behavior recorded during first cycle after zero field cooling is repeated during all subsequent cycles.

\begin{figure}[htb]
\begin{center}
\includegraphics[width=\columnwidth]{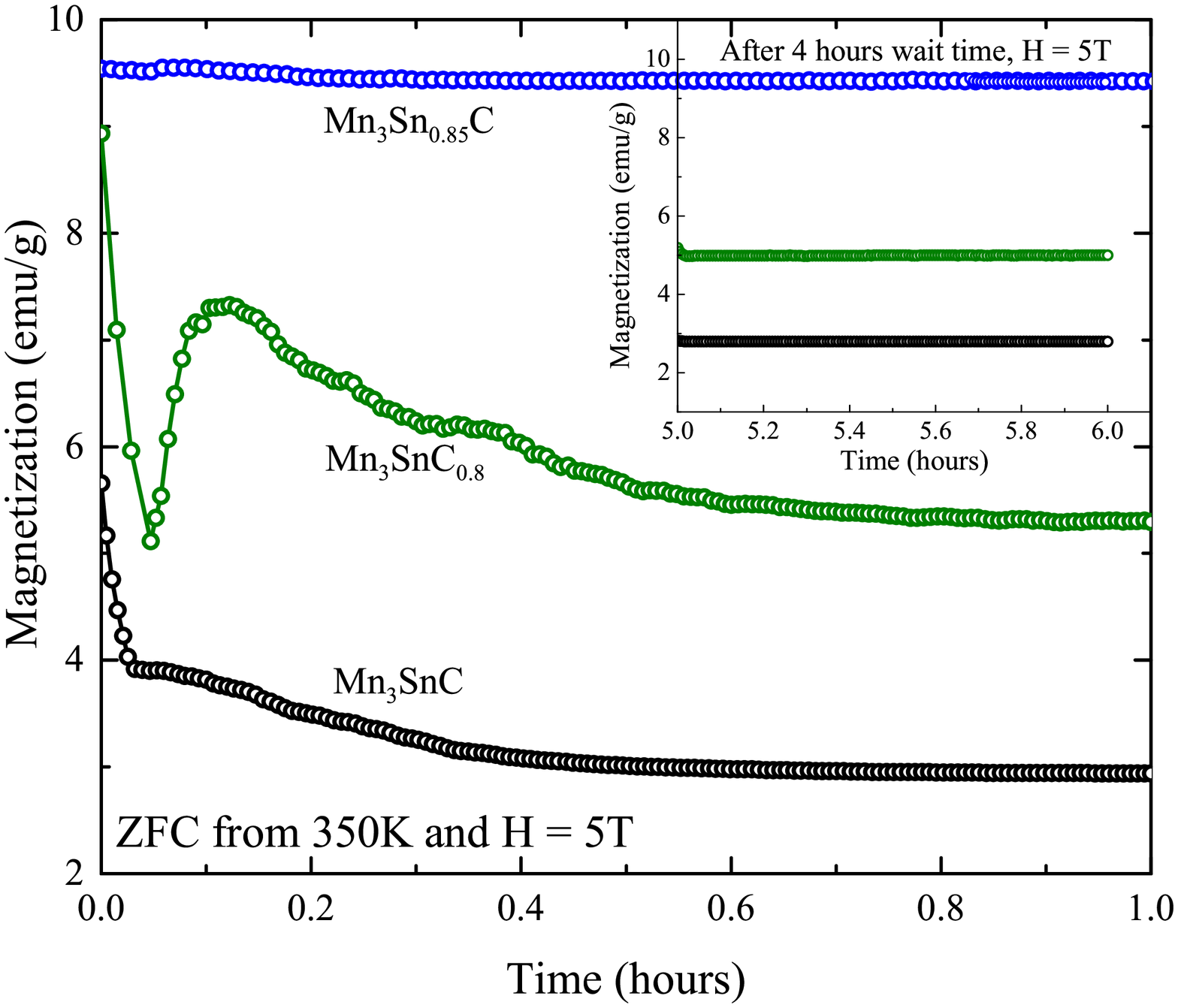}
\caption{Time dependent magnetization behavior in Mn$_3$SnC, Mn$_3$SnC$_{0.8}$ and Mn$_3$Sn$_{0.85}$C at their respective $T_{ms}$ for the first one hour after application of magnetic field, H = 5T. The samples were first cooled in zero field from 350K to $T_{ms}$. Inset shows magnetization as a function of time, five hours after the initial application of field.}
\label{fig:Mtime}
\end{center}
\end{figure}

Inverted hysteresis loops in Mn$_3$SnC have been ascribed to be due to a time delay between ferromagnetic and antiferromagnetic ordering \cite{PhysRevB.96.014436}. Therefore absence of inverted M(H) loops raises questions about the nature and interplay of ferromagnetic and antiferromagnetic interactions in Mn$_3$Sn$_{0.85}$C. While, from the isothermal magnetization it is clear that both, ferromagnetic and antiferromagnetic interactions are present in Mn$_3$Sn$_{0.85}$C, their time evolution is not very clear. To further understand the magnetization dynamics, time dependent magnetization studies were carried out on these three compounds, Mn$_3$SnC, Mn$_3$SnC$_{0.8}$ and Mn$_3$Sn$_{0.85}$C and these results are presented in Fig. \ref{fig:Mtime}.

The time dependent magnetization data presented here in Fig. \ref{fig:Mtime} though follows a similar protocol as in Ref. \cite{PhysRevB.96.014436}, have been recorded at $T_{ms}$. Here too, the samples were cooled in zero field from 350K to $T_{ms}$ and a field of 5T was applied and magnetization was recorded for one hour. The resulting magnetization shows a initial rapid decrease followed by much slower decrease. The rapid decrease has been ascribed to onset of antiferromagnetic interactions while the lattice expansion at the $T_{ms}$ is responsible for slower dynamics. A pronounced discontinuity is also visible due to equilibration of the two magnetization processes. Distinct differences emerge in the magnetization relaxation process in the three compounds. While the discontinuity, in comparison to Mn$_3$SnC, is highly pronounced in Mn$_3$SnC$_{0.8}$, it becomes very feeble in Mn$_3$Sn$_{0.85}$C. Such a behavior is not entirely surprising given the fact that both temperature dependence of magnetization as well as lattice constant shows a broad magnetostructural transition in Mn$_3$Sn$_{0.85}$C. A pronounced discontinuity as seen in case of Mn$_3$SnC and Mn$_3$SnC$_{0.80}$ implies, spin flipping from vertical (along (001) direction) to horizontal (in the $x-y$ plane) occurs at a much faster rate as compared to lattice expansion. Estimation of the two characteristic relaxation times, ($\tau_1$ and $\tau_2$) using the relation $M(t) = M(0)\exp(-t/\tau)$ supports this point. In Mn$_3$SnC, the two characteristic times were estimated as, $\tau_1$ = 75s and $\tau_2$ = 708s and in case of carbon deficient compound their values were 109s and 974s respectively. On the other hand the estimated values of $\tau_1$ and $\tau_2$ in Mn$_3$Sn$_{0.85}$C were 335s and 357s respectively. Similar values of $\tau_1$ and $\tau_2$ could be the reason for negligible discontinuity in time dependent magnetization in Mn$_3$Sn$_{0.85}$C. Further, a comparison of values of $\tau_1$ in the three compounds indicate that antiferromagnetic spin flipping slows down with both, C and Sn deficiency and the slowing down effect is much more in Mn$_3$Sn$_{0.85}$C. At the same time $\tau_2$ values indicate that, in Mn$_3$SnC$_{0.8}$ compound the lattice becomes more rigid compared to Mn$_3$SnC while Sn vacancies allow a facile expansion of lattice at the structural transition.

In the time dependent measurements, after recording magnetization of the compound as a function of time for one hour, the field was withdrawn and the sample temperature was held constant at $T_{ms}$ for about four hours. The field of 5T was reapplied and magnetization was recorded for the next one hour. These data for all the three compounds is presented in the inset of Fig. \ref{fig:Mtime}. It can be seen that the withdrawal and reapplication of magnetic field has little or no effect on the magnetic order that prevailed at the end of the first hour. This indicates that all dynamics have seized and an equilibrium situation has been established within the first one hour. This is also indicated by the values of $\tau_2$ which are about 6 minutes in Mn$_3$Sn$_{0.85}$C compound and 12 to 16 minutes in the other two compounds. Under equilibrium conditions, Mn$_3$SnC has the least ferromagnetic moment, and it increases for both, Mn$_3$SnC$_{0.8}$ and Mn$_3$Sn$_{0.85}$C. The Mn$_3$Sn$_{0.85}$C compound has the highest ferromagnetic moment at its  $T_{ms}$. This indicates that ferromagnetic interactions strengthen with the introduction of C or Sn vacancies in the antiperovskite lattice.

\begin{figure}[htb]
\begin{center}
\includegraphics[width=\columnwidth]{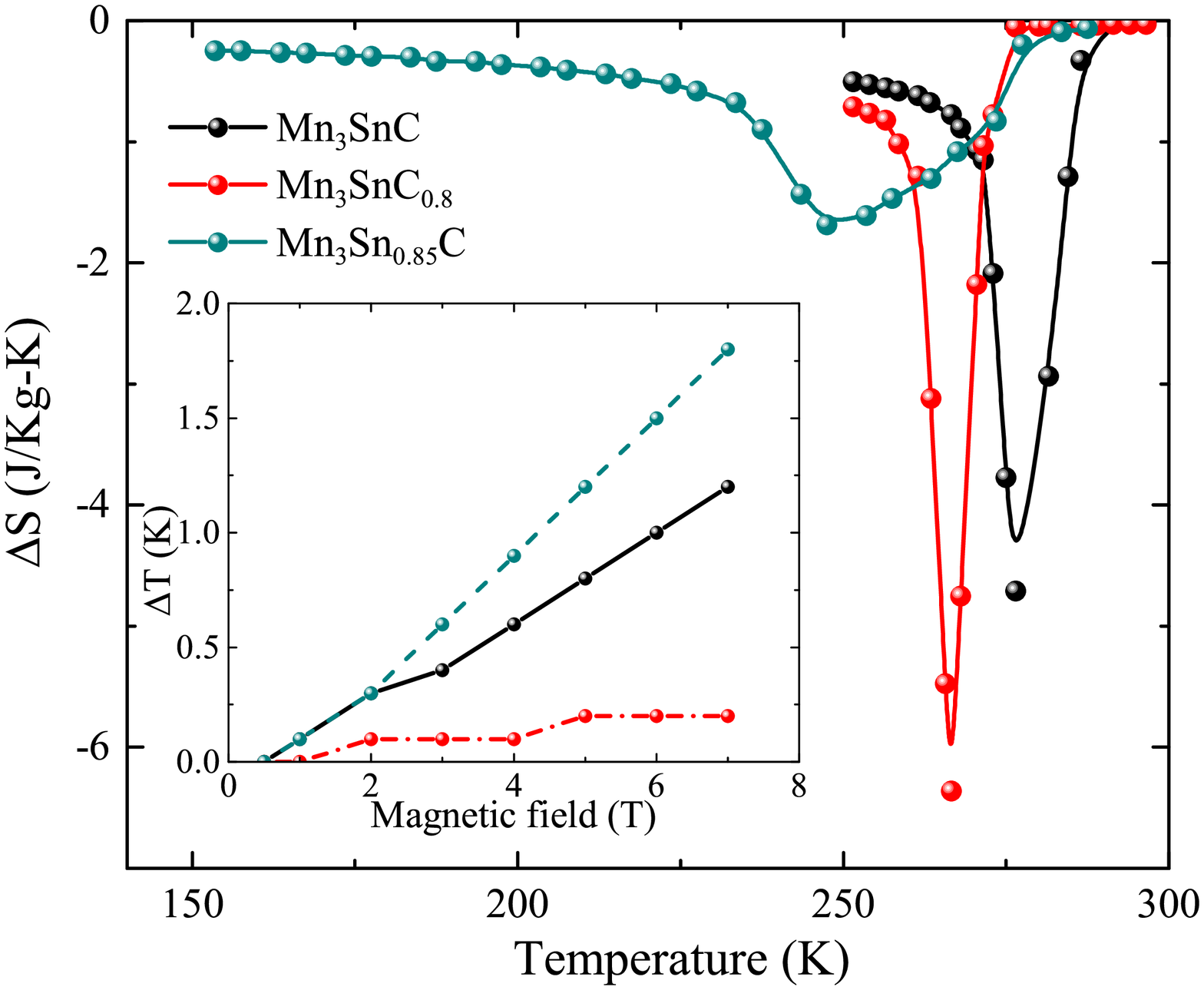}
\caption{Isothermal entropy change in Mn$_3$SnC, Mn$_3$SnC$_{0.8}$ and Mn$_3$Sn$_{0.85}$C around their respective $T_{ms}$ calculated for 0T $\le$ H $\le$ 7T. Inset shows variation of change in $\Delta T$ with magnetic field.}
\label{fig:mce}
\end{center}
\end{figure}

The nature of magnetocaloric effect (MCE) is very sensitive to type and strength of magnetic interactions. While Mn$_3$GaC exhibits a large inverse MCE due to its antiferromagnetic ordering, the presence of ferromagnetic interactions result in a normal MCE in Mn$_3$SnC. In Fig. \ref{fig:mce}, isothermal entropy changes occurring in the three compounds at their $T_{ms}$ are presented. The entropy changes are normal in all three compounds but the magnitude of entropy change is much smaller in case of Mn$_3$Sn$_{0.85}$C compared to the other two compounds. A marginal increase in entropy is noted in case of Mn$_3$SnC$_{0.8}$ as compared to Mn$_3$SnC. The decrease in magnitude of the entropy in Mn$_3$Sn$_{0.85}$C is surprising because of the presence of strong ferromagnetic interactions as indicated by time dependent studies and hence can be only reconciled to be due to presence of equally strong antiferromagnetic interactions. A competition between these two interactions can result in both, broad transition and a smaller entropy change. Further, studies on Mn$_3$Ga$_{1-x}$Sn$_x$C have shown that the strength of antiferromagnetic interactions are critically related to distortions of Mn$_6$C octahedra or lattice strain and the effect of lattice strain can be seen in the variation of entropy peak as a function of magnetic field \cite{Dias2015117}. To check the effect of lattice strain in all the three samples, the change in peak temperature of entropy peak, $\Delta T = T_{peak}(H) - T_{peak}(0.1T)$ is plotted as a function of applied field in the inset of Fig. \ref{fig:mce}. Smaller the change in peak temperature more rigid are the Mn$_6$C octahedra. From the figure, it can be clearly seen that the Mn$_6$C octahedra are most rigid in case of Mn$_3$SnC$_{0.8}$ while they experience a lesser strain in Mn$_3$Sn$_{0.85}$C. Therefore presence of both, antiferromagnetic and ferromagnetic interactions with nearly equal strength in Mn$_3$Sn$_{0.85}$C gains weight. A deficiency of carbon, however, results in more rigid Mn$_6$C octahedra and stronger ferromagnetic interactions. This scenario is also supported by the increase in magnitude of MCE peak in Mn$_3$SnC$_{0.8}$ as compared to Mn$_3$SnC.

Thus the magnetization studies on Mn$_3$SnC, Mn$_3$SnC$_{0.8}$ and Mn$_3$Sn$_{0.85}$C indicate that C and Sn do play a role in magnetostructural transformation in this antiperovskite. The primary role of Sn and C is to allow distortions in the manganese octahedra. Previous EXAFS studies on such Mn based antiperovskites have shown that such distortions result in long and short Mn-Mn bonds resulting in ferromagnetic and antiferromagnetic interactions respectively \cite{Dias4996933,Dias201548}. Particularly in case of Mn$_3$SnC, the distortions are such that one third of the Mn--Mn distances elongate while the other two third shorten with respect to the distance expected for cubic symmetry \cite{Dias201548}. This would imply, elongation of Mn$_6$C octahedra along the $z-$ axis and a contraction in the $x-y$ plane. In the present study, C deficiency results in strengthening of ferromagnetic interactions. It is possible that carbon vacancy at the centre of the octahedra, results in further elongation of the octahedra. The elongation produces a tensile strain making the octahedra more rigid as well as strengthening the ferromagnetic interactions due to longer Mn--Mn distance.  In comparison with stoichiometric Mn$_3$SnC, the effect of stronger ferromagnetic interactions and more rigid octahedra is responsible for presence of inverted magnetic hysteresis loops over a larger temperature as well magnetic field range, strong discontinuity in the time dependent magnetization measurements, increase magnitude of the entropy peak and a smaller $\Delta T$ with increasing magnetic field. On the other hand, the deficiency of Sn atoms results in alleviating the strain on Mn$_6$C octahedra and allows for a wider distribution of Mn-Mn bond distances. Such a distribution is responsible of a broad magnetic transition as well as stronger ferromagnetic and antiferromagnetic interactions as compared to Mn$_3$SnC.

\section{Conclusions}
In conclusion, the present studies on Mn$_3$SnC$_{0.8}$ and Mn$_3$Sn$_{0.85}$C antiperovskites indicate that both C and Sn atoms distort the Mn$_6$C octahedra so as to alter the dynamics as well as statics between ferromagnetic and antiferromagnetic interactions. While deficiency of C results in a tensile strain on the Mn$_6$C octahedra which strengthens the ferromagnetic interactions, Sn deficiency alleviates the strain allowing for a wider distribution of Mn--Mn bond distances resulting in strengthening of both, ferromagnetic and antiferromagnetic interactions. Thus, both Sn and C play a role of confining Mn atoms to form a distorted octahedra which is responsible for complex magnetic state seen in Mn$_3$SnC.

\section*{Acknowledgement}
Council for Scientific and Industrial Research (CSIR), New Delhi is gratefully acknowledged for financial assistance under 03(1343)/16/EMR-II. Authors thank the Department of Science and Technology, India for the travel support, Saha Institute of Nuclear Physics and Jawaharlal Nehru Centre for Advanced Scientific Research, India for facilitating the experiments at the Indian Beamline, Photon Factory, KEK, Japan. VNG thanks University Grants Commission for BSR Fellowship.

\bibliographystyle{elsarticle-num}
\bibliography{References}

\begin{thebibliography}{10}
\expandafter\ifx\csname url\endcsname\relax
  \def\url#1{\texttt{#1}}\fi
\expandafter\ifx\csname urlprefix\endcsname\relax\def\urlprefix{URL }\fi
\expandafter\ifx\csname href\endcsname\relax
  \def\href#1#2{#2} \def\path#1{#1}\fi

\bibitem{Tohei200394}
T.~Tohei, H.~Wada, T.~Kanomata, J. Appl. Phys. 94~(3) (2003) 1800--1802.

\bibitem{Lewis200393}
M.-H. Yu, L.~H. Lewis, A.~R. Moodenbaugh, J. Appl. Phys. 93~(12) (2003)
  10128--10130.

\bibitem{Aczel201490}
A.~A. Aczel, L.~Li, V.~O. Garlea, J.-Q. Yan, F.~Weickert, M.~Jaime, B.~Maiorov,
  R.~Movshovich, L.~Civale, V.~Keppens, D.~Mandrus, Phys. Rev. B 90 (2014)
  134403.

\bibitem{Kamishima200063}
K.~Kamishima, T.~Goto, H.~Nakagawa, N.~Miura, M.~Ohashi, N.~Mori, T.~Sasaki,
  T.~Kanomata, Phys. Rev. B 63 (2000) 024426.

\bibitem{Li200572}
Y.~B. Li, W.~F. Li, W.~J. Feng, Y.~Q. Zhang, Z.~D. Zhang, Phys. Rev. B 72
  (2005) 024411.

\bibitem{Zhang2014115}
X.~H. Zhang, Y.~Yin, Q.~Yuan, J.~C. Han, Z.~H. Zhang, J.~K. Jian, J.~G. Zhao,
  B.~Song, J. Appl. Phys. 115~(12) (2014) 123905.

\bibitem{Wu2013114}
M.~Wu, C.~Wang, Y.~Sun, L.~Chu, J.~Yan, D.~Chen, Q.~Huang, J.~W. Lynn, J. Appl.
  Phys. 114~(12) (2013) 123902.

\bibitem{Chi2001120}
E.~O. Chi, W.~S. Kim, N.~H. Hur, Solid State Commun. 120~(7-8) (2001) 307--310.

\bibitem{Sun201062}
Y.~Sun, C.~Wang, L.~Chu, Y.~Wen, M.~Nie, F.~Liu, Scripta Mater. \textbf{62}~(9)
  (2010) 686--689.

\bibitem{Shibayama2011109}
T.~Shibayama, K.~Takenaka, J. Appl. Phys. \textbf{109}~(7) (2011) 07A928.

\bibitem{Lukashev200878}
P.~Lukashev, R.~F. Sabirianov, K.~Belashchenko, Phys. Rev. B \textbf{78} (2008)
  184414.

\bibitem{Wang200985}
B.~S. Wang, P.~Tong, Y.~P. Sun, X.~Luo, X.~B. Zhu, G.~Li, X.~D. Zhu, S.~B.
  Zhang, Z.~R. Yang, W.~H. Song, J.~M. Dai, Europhys. Lett. 85~(4) (2009)
  47004.

\bibitem{Dias201548}
E.~T. Dias, K.~R. Priolkar, A.~Das, G.~Aquilanti, $\rm\ddot{O}$.
  $\rm\c{C}$akir, M.~Acet, A.~K. Nigam, J. Phys. D: Appl. Phys.
  \textbf{48}~(29) (2015) 295001.

\bibitem{Fruchart19708}
D.~Fruchart, E.~Bertaut, F.~Sayetat, M.~N. Eddine, R.~Fruchart, J.~Sénateur,
  Solid State Commun. 8~(2) (1970) 91--99.

\bibitem{Cakir2014115}
$\rm\ddot{O}$. $\rm\c{C}$akir, M.~Acet, M.~Farle, A.~Senyshyn, J. Appl. Phys.
  \textbf{115}~(4) (2014) 043913.

\bibitem{PhysRevB.96.014436}
O.~\ifmmode \mbox{\c{C}}\else \c{C}\fi{}ak\ifmmode \imath \else~\i \fi{}r,
  F.~Cugini, M.~Solzi, K.~Priolkar, M.~Acet, M.~Farle, Phys. Rev. B 96 (2017)
  014436.

\bibitem{Dias4996933}
E.~T. Dias, K.~R. Priolkar, R.~Ranjan, A.~K. Nigam, S.~Emura, Journal of
  Applied Physics 122~(10) (2017) 103906.

\bibitem{Dias2014363}
E.~Dias, K.~Priolkar, A.~Nigam, J. Magn. Magn. Mater. 363~(0) (2014) 140 --
  144.

\bibitem{Dias2015117}
E.~T. Dias, K.~R. Priolkar, $\rm\ddot{O}$. $\rm\c{C}$akir., M.~Acet, A.~K.
  Nigam, J. Appl. Phys. 117~(12) (2015) 123901.

\end{thebibliography}

\end{document}